\documentclass{article}
\usepackage{graphicx}
\newcommand{\B}{\mathbf{B}}
\newcommand{\C}{\mathbf{C}}
\newcommand{\W}{\mathbf{W}}
\newcommand{\A}{\mathbf{A}}
\newcommand{\M}{\mathbf{M}}
\newcommand{\ess}{\mathbf{S}}
\newcommand{\I}{\mathbf{I}}
\newcommand{\x}{\mathbf{x}}

\newcommand{\uu}{\mathbf{u}}
\newcommand{\vv}{\mathbf{v}}

\newcommand{\PP}{\mathbf{P}}
\newcommand{\LL}{\lambda}
\newcommand{\lb}{\langle}
\newcommand{\rb}{\rangle}

\newcommand{\Lmat}{\mbox{\boldmath$\Lambda$}}
\renewcommand{\baselinestretch}{1.5}
\begin{document}
\title{Performance Variability and Project Dynamics}%
\author{Bernardo A. Huberman and Dennis M. Wilkinson
\thanks{HP Labs, 1501 Page Mill Rd., Palo Alto, CA, 94304}}%
%\address{HP Labs, 1501 Page Mill Rd., Palo Alto, CA, 94304}%
%\email{}%
%\thanks{}%
%\subjclass{}%
%\keywords{}%
%\date{}%
%\dedicatory{}%
%\commby{}%
% ----------------------------------------------------------------
\maketitle
\begin{abstract}
We present a dynamical theory of complex cooperative projects such
as large engineering design or software development efforts,
comprised of concurrent and interrelated tasks. The model accounts
for temporal fluctuations both in task performance and in the
interactions between related tasks. We show that as the system
size increases, so does the average completion time. Also, for
fixed system size, the dynamics of individual project realizations
can exhibit large deviations from the average when fluctuations
increase past a threshold, causing long delays in completion
times. This effect is in agreement with empirical observations,
and can be mitigated by arranging projects in a hierarchical or
modular structure.
\end{abstract}
%\maketitle

% ----------------------------------------------------------------
\newpage
\section{Introduction}
\label{s:introduction} One of the main challenges in a large
design project or, more generally, any problem solving process is
the coordination of the different tasks comprising that project or
process
\cite{Galbraith_org,Thompson_org,Huberman_Hogg_cooperation}.
Coordination is particularly difficult in situations where the
complexity of the project leads to its division into concurrent,
interdependent tasks whose results must be dynamically integrated
into an overall satisfactory solution. Examples are large-scale
software design projects, engineering design projects, and
industrial research and development efforts. It has long been
recognized that because of the coupled nature of the component
tasks such problem solving processes are inherently iterative in
their execution \cite{Kline_org,Whitney_org}. Information from the
partial solution to a given task can trigger a chain of revisions
as solutions to other, related tasks are modified
\cite{Allen_1966, Clark_Fujimoto, Thomke_1997, Van_Zandt}.

A number of formal models of cooperative processes have been
proposed that predict the number of iterations required for
completion, while suggesting optimal concurrency and iteration
schemes which minimize the overall time or cost (e.g.,
\cite{Ahmadi_Wang,Ha_Porteus,Krishnan_et_al_1997,Jin_Levitt,Ford_Sterman_1998,Roemer_et_al_2000,Loch_et_al_2001,Pena_Mora_Li}).
One example of these approaches is the work transformation matrix
(WTM) model \cite{Smith_Eppinger}, which is an extension of the
design structure matrix model (see \cite{Browning_2001_review} for
a review). It provides a simple mathematical representation of
design iterations that estimates the number of iterations and work
required for a given arrangement of coupled tasks.

Helpful as these models are, they do not account for the frequent
situation in which projects enter a vicious cycle of continuing
revisions \cite{Cusumano_Selby,Joglekar_2001,Terwiesch_et_al_2002}
resulting in budget overruns, missed business opportunities and
delays which, at times, can make the final solution obsolete
\cite{Brooks,Morris_Hugh,Terwiesch_Loch_1999}. A major factor
leading to these undesirable outcomes stems from the unpredictable
fluctuations in the value that a particular unit of work on a
single task brings to the overall project. In order to take into
account these fluctuations, recent models have focused on
individual sources of variability, including asynchronicity and
random timing of task updates and information exchange
\cite{MihmLochHuchzermeier,Huberman_Glance_continuous},
information withholding \cite{Yassine_et_al}, exogenous changes
\cite{Japanese_software,Mar_1999}, volatility of resource
allocation \cite{Repenning_et_al_2001}, behavioral choice
\cite{Ford_Sterman_1999}, uncertainty of performance evaluation
\cite{Browning_et_al_2002}, and the complicated landscape of
performance maxima and minima \cite{MihmLochHuchzermeier}.
However, since these models are not stochastic in nature, they do
not account for the random and unpredictable nature of project
dynamics, which is an intrinsic and unavoidable element of any
complex cooperative project.

In this paper, we present a mathematical model of iterative
problem solving processes which explicitly incorporates the
fluctuating nature of task performance and interdependence. The
model uses the formulation of the work transformation matrix to
describe the dynamics of the project in terms of the distance to
solution of the component tasks. The crucial fluctuating component
is included in each updating step. Progress towards an overall
solution is thus represented by a stochastic dynamic process which
captures the erratic nature of the evolution of real-world
projects and their component tasks. We first show that the time to
solution increases on average as the number of interactions in the
project increase, in agreement with empirical results
\cite{Clark_1989,Griffin_1997,Reel_1999,Japanese_software} and
previous simulations \cite{MihmLochHuchzermeier,Yassine_et_al}. We
also show that as the average strength of the fluctuations
increases, the time to completion increases, also in agreement
with empirical studies \cite{Ibbs}. We demonstrate that in a large
project comprising many interactive tasks, a hierarchical or
modular structure can, on average, alleviate the problem of large
convergence time, as previously proposed (e.g. \cite{Hammer_1996})
and in agreement with empirical results
\cite{Ulrich_1995,Sosa_et_al_2003} and previous studies
\cite{MihmLoch_andus,Rivken_Siggelkow,Ethiraj_Levinthal}.

Moreover, because of the temporal variability of the fluctuations,
evolution towards solution of any particular instance of a project
can differ greatly from the average behavior. As a result,
projects which would converge smoothly to solution in the absence
of fluctuations  can deviate significantly from this path. When
the temporal variability is low, convergence to a solution is
smooth and the finishing times are normally distributed close to
the average value. But above a given threshold in variability, the
distribution of finishing times undergoes a transition to a
heavy-tailed log-normal form. This distribution is in agreement
with earlier empirical results on cooperative problem solving
\cite{Clearwater_science} and implies possible finishing times far
greater than the average.

Finally, we show that the effect of fluctuations is more severe in
projects which converge slowly to solution. This implies that
hierarchical organization not only can decrease the average time
to solution, but can also mitigate the possibly unavoidable
effects of fluctuating interactions.

The paper is organized as follows. In \S \ref{s:model} we present
the dynamical model and generalize it so as to take into account
both synchronous and asynchronous updates of the problem solving
process. In \S \ref{s:avg_dyn} we study the model's average
behavior and the relation between convergence time and project
size.  In \S \ref{s:fluct} we consider the effects of fluctuating
efforts and establish regimes that lead to large deviations away
from the smooth convergence to the goal. Section
\ref{s:conclusion} summarizes our results.

\section{A Model of Group Problem Solving} \label{s:model}
Consider a group of individuals, or even computer programs,
working cooperatively towards the solution of a problem. An
example of this process in the design space is provided by a large
software development effort in which individuals or teams work
iteratively on pieces of the software and pass it along to others
so that whole modules and eventually the whole application may be
completed. Alternatively, one can conceive of teams of engineers
involved in the design of a complicated mechanism that requires
tight integration of all the parts in order to achieve the desired
goal.

The progress of a cooperative project comprising $n$ tasks towards
a solution can be represented mathematically  by  a state vector
$\x$ with $n$ components, as in the WTM model of Smith and
Eppinger \cite{Smith_Eppinger}. The $i$th element of $\x$
represents how far task $i$ is from completion. The dynamics of
the system are specified by a time-dependent interaction matrix
$\A_t$ which encapsulates the interdependencies among the tasks.
The state of the overall project at time $t+1$ is determined by
\begin{equation}\label{dynamics}
 \x_{t+1} = \A_t\x_t
\end{equation}
where $\x_t$ is the state at time $t$. The initial state $\x_1$
may be taken to be a vector of 1's, defining the scale by which we
measure how much work is left on the tasks. Ideally, as the
project proceeds, the entries of the state vector will become
smaller and smaller as the tasks near completion.

The interpretation of the elements of the interaction matrix is as
follows (see also the example below). The off-diagonal elements
measure how much one task's partial solution helps or hinders the
completion of the other tasks. A positive entry signifies that one
unit of work on task $j$ causes $(A_t)_{ij}$ units of rework on
task $i$, while a zero entry means that tasks $i$ and $j$ have no
direct effect on each other. In the WTM model, only positive or
zero entries were allowed. Our theory allows for negative entries
when efforts on task $j$'s hasten the completion of task $i$. The
diagonal elements of the interaction matrix $\A_t$ account for
different rates of progress on different tasks. This is in
contrast to the WTM model where all tasks are performed at the
same rate.

As an example, consider the following hypothetical development of
an internet-based software client. The process has been divided
into three modules: user interface, database front end, and
network layering. It is thus modelled by a three-dimensional
system as follows:
\[
\x_t = \left(
\begin{array}{c}
\mbox{work remaining on user interface} \\
\mbox{work remaining on database front end} \\
\mbox{work remaining on network layering} \\
\end{array}
\right)
\]
The initial state $\x_0$ of the system is a vector of 1's and the
time step is one week. After a week, the state of the system is
given by
\begin{equation} \label{example1week}
\x_2 = \A_1 \x_1 = \left(
\begin{array}{ccc}
 0.65 & 0.2 & 0.05 \\
 0.3 & 0.6 & 0.07 \\
 0.09 & 0.05 & 0.7 \\
\end{array}
\right) \left(
\begin{array}{c}
1 \\
1 \\
1 \\
\end{array}
\right) = \left(
\begin{array}{c}
0.9 \\
0.97 \\
0.84 \\
\end{array}
\right)
\end{equation}
where $\A_1$ is a hypothetical interaction matrix which describes
the dynamics over the first week of the project.

Note that in this example the interdependencies between tasks slow
down the overall progress. The tasks proceed rather quickly when
taken alone (diagonal elements), but due to the strong
interdependencies (off-diagonal elements), not much progress is
made overall in the first week. The off-diagonal elements in this
example were chosen to reflect the difficulties in making
components of a large software effort compatible.

\subsubsection*{Project architecture} The matrices $\A_t$ entirely
define the dynamics of the system by representing the interactions
between tasks (off-diagonal elements) and the task work rates
(diagonal elements). To a large degree, properties of the
project's architecture define the elements of the interaction
matrices. Such properties include the inherent interdependence of
related tasks, the structure of the organization (hierarchical,
flat, etc.), and the relationships between individuals working on
tasks and communities of practice
\cite{Huberman_Hogg_communities_practice}, among others. Unless
the project undergoes a major reorganization\footnote{In the case
of project reorganizations, we would introduce a second
unperturbed WTM to account for the reorganization. The first part
of the project would be modelled using the first $\A_0$ plus
fluctuations, and the second part using the second $\A_0$ plus
fluctuations. This reorganization may thus be easily accounted for
and will not produce any interesting new dynamics beyond what is
discussed in this paper.}, these interactions remain constant over
the project's lifetime.

Interdependencies of this type give a constant, underlying
structure to the system's dynamics and form a basic interaction
matrix $\A_0$. This matrix is simply the work transformation
matrix \cite{Smith_Eppinger}, which we will refer to as the
\textit{unperturbed} interaction matrix or unperturbed WTM. The
actual interaction matrices $\A_t$ in our model are created by
introducing fluctuations into the elements of the unperturbed WTM.
In practice, an unperturbed WTM can be created for a particular
project by assigning numerical values to the project's design
structure matrix (DSM). The DSM is an important tool (see
\cite{Browning_2001_review} for a review) which encapsulates a
project's interdependencies, usually by means of a survey of the
individuals involved. For example, McDaniel \cite{McDaniel_1996}
created a WTM for the appearance design of a vehicle's interior by
assigning each qualitative DSM entry a numerical value and
averaging the values obtained from many sources. This process is
shown in tables \ref{t:DSM} and \ref{t:WTM}. The numerical values
chosen for the interdependencies are of crucial importance to the
system's evolution, as explained in sections \ref{s:avg_dyn} and
\ref{s:fluct}.
\renewcommand{\baselinestretch}{1}
\begin{table} [h!]
\centering \footnotesize
\renewcommand{\baselinestretch}{1}
\begin{tabular}{l|c|c|c|c|c|c|c|c|c|c|}
 & 1 & 2 & 3 & 4 & 5 & 6 & 7 & 8 & 9 & 10 \\ \hline
 1. Carpet &  & W & 0 & W & W & 0 & 0 & 0 & W & 0 \\
 2. Center console & W &  & W & 0 & 0 & S & W&0&M&0\\
 3. Door trim panel & 0&W& &W&0&M&0&0&M&W \\
 4. Garnish trim &W&0&M& &0&W&W&0&W&0\\
 5. Overhead system &W&0&0&0& &0&0&0&0&0\\
 6. Instrument panel &0&S&S&W&0& &W&0&0&M\\
 7. Luggage trim &0&0&0&W&0&W& &W&W&0\\
 8. Package tray &0&0&0&W&0&0&W& &M&0\\
 9. Seats &W&M&M&W&0&W&W&M& &M\\
 10. Steering wheel &0&0&0&0&0&S&0&0&M& \\ \hline
 \end{tabular}
\caption{Sample DSM characterizing interdependencies in appearance
design of vehicle interior. Here S means strong, M medium, W weak
and 0 none.} \label{t:DSM}
\end{table}
\begin{table} \label{t:WTM}
\centering \footnotesize
\begin{tabular}{l|c|c|c|c|c|c|c|c|c|c|}
 & 1 & 2 & 3 & 4 & 5 & 6 & 7 & 8 & 9 & 10 \\ \hline
 1. Carpet & 0.85 & 0.06 & 0.01 & 0.03 & 0.03 & 0 & 0 & 0 & 0.03 & 0 \\
 2. Center console & 0.05 & 0.53 & 0.02 & 0 & 0 & 0.15 & 0.01 &0&0.12&0.01\\
 3. Door trim panel & 0.01&0.02&0.47 &0.04&0&0.12&0.01&0&0.09&0.01 \\
 4. Garnish trim &0.03&0&0.09&0.68&0&0.07&0.05&0.01&0.04&0\\
 5. Overhead system &0.02&0&0&0&0.83 &0&0&0&0&0\\
 6. Instrument panel &0&0.15&0.13&0.08&0&0.28&0.03&0&0.01&0.1\\
 7. Luggage trim &0&0.01&0.01&0.05&0&0.03&0.76&0.03&0.04&0\\
 8. Package tray &0&0&0&0.05&0&0&0.03&0.83&0.08&0\\
 9. Seats &0.04&0.12&0.09&0.04&0&0.02&0.02&0.08&0.63 &0.1\\
 10. Steering wheel &0&0.01&0.01&0&0&0.13&0&0&0.1&0.7 \\ \hline
 \end{tabular}
\caption{WTM for vehicle interior appearance design process
\cite{McDaniel_1996} created from averaging many DSM's like the
one shown in table \ref{t:DSM}. Here S interdependencies were
given the value 0.15, M 0.1, and W 0.05. The diagonal elements
were obtained by estimating the autonomous completion time for
each component.}
\end{table}

\subsection*{Fluctuations} Interactions in a cooperative problem
solving process do not remain constant on short time scales but
vary due to a number of features, including asynchronicity of
information exchange, exogenous changes, behavioral choice,
uncertainty of performance evaluation, the varying relevance of
the partial solutions of related tasks to each other, and other
factors as discussed in the introduction. These factors conspire
to produce fluctuations in the elements of the interaction matrix.
The exact nature of the fluctuations in any project is impossible
to determine in advance because of the complexity of the process
and of these interacting factors. The problem solving effort is
thus effectively a stochastic process, where the fluctuations
introduce a random component into the dynamics.

To see how these stochastic fluctuations affect the dynamics,
consider the software development example introduced above. The
evolution of the project in the first week was determined as shown
in equation (\ref{example1week}). During the second week, a
different story unfolds, as the fluctuations alter the entries of
the interaction matrix :
\[
\x_3 = \A_2 \x_2 = \left(
\begin{array}{ccc}
 0.8 & 0.27 & -0.05 \\
 0.23 & 0.55 & 0.04 \\
 -0.03 & 0.02 & 0.7 \\
\end{array}
\right) \left(
\begin{array}{c}
0.9 \\
0.97 \\
0.8 \\
\end{array}
\right) = \left(
\begin{array}{c}
0.9419 \\
0.7725 \\
0.5524 \\
\end{array}
\right)
\]

Note in particular that the interaction matrix has changed, so
that the state vector changes in a different way from its
evolution during the first week. During the second week, progress
was slow on the user interface component (interaction matrix
element (1,1)). Moreover, hindering interactions, perhaps due to a
lack of communication with the front end developer (elements (1,2)
and (2,1)), resulted in lost ground on this task. The front end
itself was hurt by the bad communication but programmer 2 worked
hard (element (2,2)) and made progress. The network protocol task
also made good progress as its results were fortuitously
compatible with those of task 1 (elements (1,3), (3,1)).

Fluctuations are reflected in our model by having the elements of
the interaction matrix vary randomly in time with a given
distribution. For simplicity, we assume that the distribution does
not vary over the lifetime of the project. It is straightforward
to relax this assumption if the project undergoes a major
reorganization as explained above.

To focus on the effect of the fluctuations, we introduce new
notation for the interaction matrices $\A_t$. Recall that the
off-diagonal elements of $\A_t$ indicate how much one unit of work
on a certain task helps or hinders a different task at time step
$t$, while the diagonal elements represent the work rate on
individual tasks. Consider a particular element of $\A_t$. One
contribution to this element comes from the unperturbed work
transformation matrix (recall the abbreviation WTM), which
represents some \textit{a priori} prediction of the value of the
interaction. The other contribution is from the unforeseen
fluctuations. This split is summarized by:
\[
%\footnotesize \centering
\mbox{\footnotesize \begin{tabular}{c}
each element of \\
interaction matrix
\end{tabular}}
=
\mbox{\footnotesize \begin{tabular}{c}
unperturbed WTM \\
(constant matrix)
\end{tabular}}
+
\mbox{\footnotesize \begin{tabular}{c}
fluctuation part \\
(matrix which varies in time)
\end{tabular}}
\]
The fluctuating part of the interaction matrix is itself a matrix
of random variables that vary in time with certain means and
variances. For each element, the fluctuation mean indicates how
far the value of the interaction is on average from the
unperturbed WTM value, and the variance indicates how much
temporal deviation there is from this mean. Since the fluctuation
means are constant throughout the project's lifetime, we can
further split the fluctuation part as follows:
\begin{eqnarray*}
\mbox{\footnotesize \begin{tabular}{c}
each element of \\
interaction matrix
\end{tabular}}
&=&
\mbox{\footnotesize \begin{tabular}{c}
unperturbed WTM \\
(constant matrix)
\end{tabular}}
+
\mbox{\footnotesize \begin{tabular}{c}
fluctuation mean \\
(constant matrix)
\end{tabular}}
+ \\
& & ~~~~~~~~~~~~ + \mbox{\footnotesize \begin{tabular}{c}
varying part of fluctuation \\
(matrix of mean-zero random variables)
\end{tabular}}
\end{eqnarray*}
The mathematical representation of this relation is
\begin{equation}\label{ABC}
  \A_t = \A_0 + (\M + \ess_t),
\end{equation}
where $\A_0$ is the unperturbed work transformation matrix, $\M$
is the matrix of mean values of the fluctuations, and $\ess_t$ is
a time-dependent matrix of mean-zero random variables representing
the temporal variation of the fluctuations. Using this notation
the dynamics of the overall problem solving process are expressed
by the stochastic equation
\begin{equation}\label{maindiscrete}
  \x_{t+1} = (\A_0 + \M + \ess_t) \x_t.
\end{equation}

The state vector $\x$ is a vector of random variables, and the
dynamics of the problem solving process are characterized by the
system's average value $\lb \x_t \rb$ and its moments $\lb
|\x_t|^p \rb$. The moments provide a measure of how likely the
system is to be found far from its average behavior.

\subsubsection*{Stopping criterion} 
\begin{figure}[h!]
     %\centering
     \includegraphics[width=2.75in]{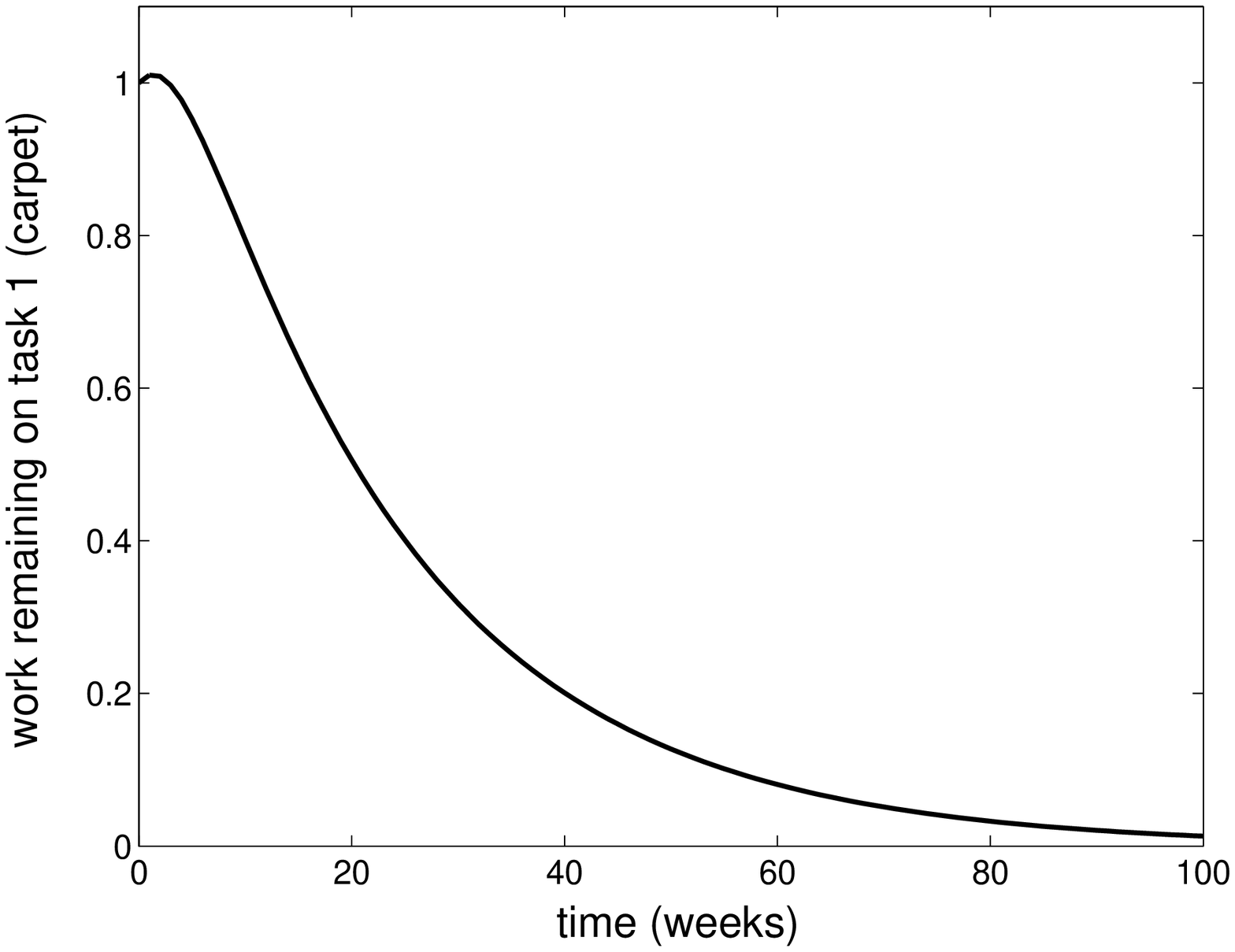}
     \includegraphics[width=2.75in]{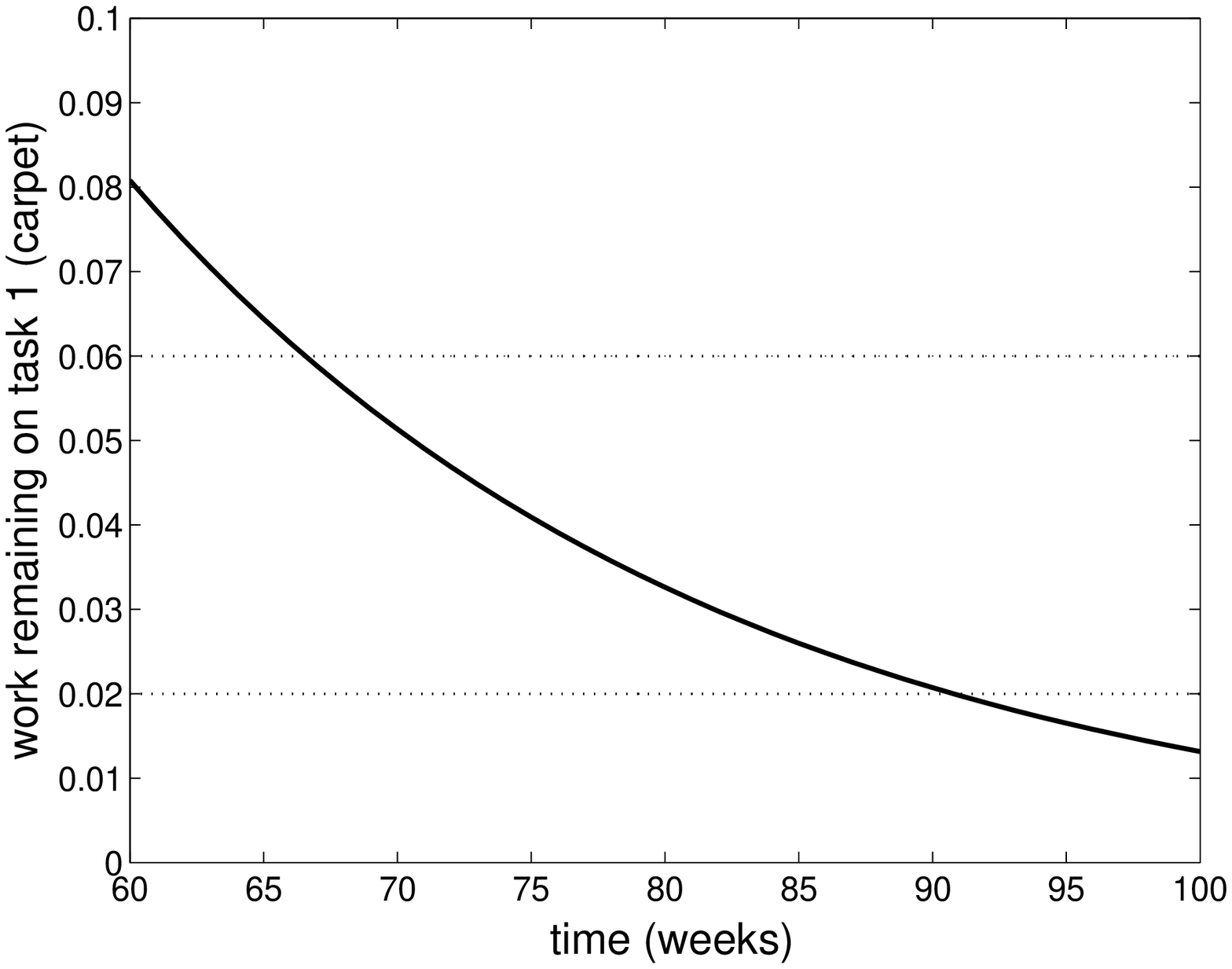}
     \caption{Simulation of project dynamics showing the
     evolution of the first component of the state vector. This plot
     was created using only the unperturbed interaction matrix $\A_0$
     shown in table \ref{t:WTM} (no fluctuations). At left, evolution over 100 weeks;
     at right, a plot of weeks 60--100 showing the difference in completion time
     for two different stopping
     criteria (dotted lines).}
    \label{f:asympt}
\end{figure}

A final important feature of
our model is the stopping criterion and the interpretation of the
zero vector. We consider the zero vector to represent an
``optimal'' solution, and the values of the state vector as an
abstract measure of the amount of work left to be done before a
task's solution is optimal. In the evolution described by equation
(\ref{dynamics}), even when $\A_t$ is constant and the dynamics
are convergent, the zero vector is never reached, as the state
vector asymptotically approaches zero as shown in figure
\ref{f:asympt}.
In a given project, the goal is thus reached when managers decide
that the tasks' values are close enough to zero. This is in
contrast to the WTM formulation in which the zero vector must be
reached for work to stop.

Our model thus implicitly contains a stopping criterion that
determines when the work is finished. For example, the stopping
criterion may be that all the elements of $\x$ are less than 0.05.
Other, task-specific criteria are of course possible.  As shown in
the right panel of figure \ref{f:asympt}, a slight change in the
stopping criterion can mean a difference of weeks or months of in
the completion time.  This interpretation is quite reasonable when
one considers that real-world projects are always terminated by
managers when the solution is deemed satisfactory by some
subjective indicators.

\subsubsection*{Continuous time model} The model proposed above is
discrete in time, implying that all the task states are updated at
once. This is applicable to situations where individuals update
their information synchronously, such as at a weekly meeting.
Conversely, in situations where information is continuously passed
between various individuals or subtasks in asynchronous fashion, a
continuous-time model is needed \cite{Huberman_Glance_continuous}.
In this continuum limit one obtains the stochastic differential
equation \cite{Oksendal}
\begin{equation}\label{maincontinuous}
  d\x = [(\A_0 + \M - \I) dt  + d\ess(t)]\x(t).
\end{equation}
where $d\ess = \ess(t) dt$ is a matrix of independent mean-zero
Wiener processes with standard deviation proportional to
$\sqrt{dt}$.

\bigskip

The introduction of fluctuations into the interaction matrix means
that our model requires few assumptions about the nature of the
project. The one assumption that is necessary, just as in the
original WTM treatment, is that the project's evolution can be
described using a linear model. This is quite reasonable because
in the early stages of a project progress on a given task is of a
general nature and can be strongly interrelated to other tasks. As
the project progresses and nears completion, partial solutions no
longer provide much new information and the effects of a given
task on others diminishes. In other words, interdependencies are
stronger at earlier stages of the problem than at later ones. The
mathematical expression of this weakening of the interdependencies
is that the interactions are proportional to the amount of work
remaining. This is a generalization of the WTM assumption that
rework caused by task correlation is proportional to the work done
on the previous stage.

\section{Average dynamics and project size} \label{s:avg_dyn}
Equations (\ref{maindiscrete}) and (\ref{maincontinuous}) describe
the evolution of the cooperative solving process towards a
solution. As discussed above, these equations are stochastic and
describe a range of possible dynamics. In order to elucidate the
possible behaviors, we first study the average dynamics of the
process. As we show later, when the fluctuations in interaction
strength are small, the evolution of the problem solving process
is very likely to be close to the average. For larger noise,
individual realizations of a given project may differ greatly from
the average. Nevertheless, this quantity is important in studying
the dependence on project organization, system size and
interaction strengths.

\begin{figure}[h!]
     \centering
      \includegraphics[width=3in]{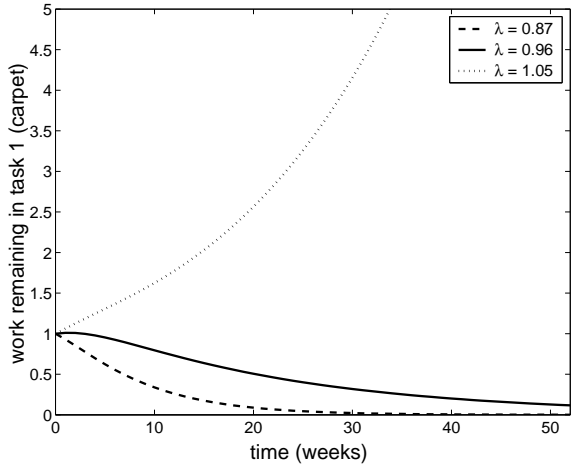}
     \caption{Simulation of average dynamics of processes with
     discrete dynamics
     and various values of $\LL$. This plot was created using the
     DSM interaction matrix for the appearance design
     of a vehicle interior (table \ref{t:DSM})
     by simply varying the numerical values
     assigned for strong, medium, and weak dependencies. The diagonal
     entries reflect the autonomous task completion rate and
     are those of the WTM in table \ref{t:WTM}.
     Here $\M = \mathbf{0}$.}
    \label{f:avg_dyn}
\end{figure}

As shown in appendix \ref{s:avg_fluct_noeffect}, the matrices
$\ess_t$ do not affect the average dynamics, so the average state
of the process evolves according to
\begin{equation}\label{avg_dyn}
  \begin{array}{cclll}
\lb \x_{t+1} \rb &=& (\A_0 + \M) \lb \x_t \rb & , & \mbox{discrete} \\
d\lb \x \rb &=&  (\A_0 + \M - \I) \lb \x(t) \rb dt & , &
\mbox{continuous}.
\end{array}
\end{equation}
This is a simple linear system whose dynamics are well-understood.
The convergence properties are completely determined by the
magnitude of the largest eigenvalue $\LL$ of the average
interaction matrix $\A_0 + \M$. That is, if $\LL < 1$, the process
will produce a satisfactory solution as the state vector $\x$
converges asymptotically to the zero vector; but if $\LL > 1$, the
process will not yield a solution as the state vector diverges, or
continues to increase. This behavior is demonstrated in figure
\ref{f:avg_dyn}.

Moreover, the value of $\LL$ determines the rate of convergence
(or divergence), with higher values of $\LL$ implying slower
convergence (or faster divergence). This property is explained
below and is demonstrated in figure \ref{f:avg_dyn}. This is the
reason why the numerical values chosen to represent the
qualitative entries of the DSM are so important, as demonstrated
in the figure.

The above convergence properties are a fundamental result of
linear algebra \cite{Strang} and have been noted in connection
with the WTM model \cite{Smith_Eppinger}. However, previous WTM
treatments have not fully explored the elegant simplicity of this
linear system, for which the asymptotic solution can be concisely
expressed. To do this, we will assume that $\A_0 + \M$ has a
simple, dominant eigenvalue\footnote{This condition is fulfilled
by almost every reasonable interaction matrix. A simple eigenvalue
has only one associated eigenvector, and thus its eigenspace is
one-dimensional. By dominant we mean that $\LL > |\LL_i|$ for all
$\LL_1 \neq \LL$. A dominant eigenvalue is necessarily real, and
the dominant eigenvalue of an interaction matrix must be positive.
There do of course exist matrices whose largest eigenvalue is not
simple, as well as matrices with multiple largest eigenvalues,
i.e. $\LL_1, \LL_2, \ldots \LL_h$ such that $|\LL_1| = |\LL_2| =
\ldots = |\LL_h|$. However, it can be easily shown (\cite{Minc},
e.g., or see \cite{Wilkinson} for an explicit discussion) that a
nonnegative system with several largest eigenvalues splits into
independent subsystems each with a simple dominant eigenvalue of
equal or smaller magnitude. If some of the elements of $\A_0 + \M$
are negative, then there can be a dominant complex conjugate pair
of eigenvalues. In this case the asymptotic behavior is not
smooth, but oscillatory. However, the convergence properties are
the same with $\LL$ replaced by the complex absolute value
$|\LL|$.}. Doing so we obtain (see appendix \ref{s:lin_dyn} for a
derivation) the following simple expression for the long-time
solution to equations (\ref{avg_dyn}):
\begin{equation}\label{avg_dyn_asympt}
  \begin{array}{cclll}
\lb \x_t \rb &=& c \uu \LL^t & , & \mbox{discrete} \\
\lb \x(t) \rb &=& c \uu e^{(\LL-1) t} & , & \mbox{continuous}
\end{array}
\end{equation}
where the constant $c$ is given by $c = \vv \cdot
\x_0$,\footnote{The case where $\vv$ is perpendicular to $\x_0$ is
irrelevant because the values of $\A_0 + \M$ are impossible to
exactly determine. In addition, we note that it is possible for
elements of $\vv$ to be very large; this situation corresponds to
matrices whose eigenvalues are highly susceptible to fluctuations.
While of great interest, such matrices are rare and usually not
reasonable interaction matrices. This topic is quite complicated
and beyond the scope of this paper; see \cite{Golubbook}, or
\cite{Wilkinson} for a discussion in relation to this particular
problem.} and $\uu$ and $\vv$ are the right and left eigenvectors,
respectively, of the matrix $\A_0 + \M$ corresponding to the
eigenvalue $\LL$. Expressions (\ref{avg_dyn_asympt}) clearly show
that the convergence rate is determined by $\LL$. It is important
to note that the long time limit may have different meanings for
different systems. If the second largest eigenvalue is very small
compared to $\LL$, the above asymptotic limit may be reached very
quickly, e.g. after as few as 2 or 3 iterations, as demonstrated
in figure \ref{f:avg_dyn} above.

\begin{figure}[h!]
     \centering
     \includegraphics[width=3in]{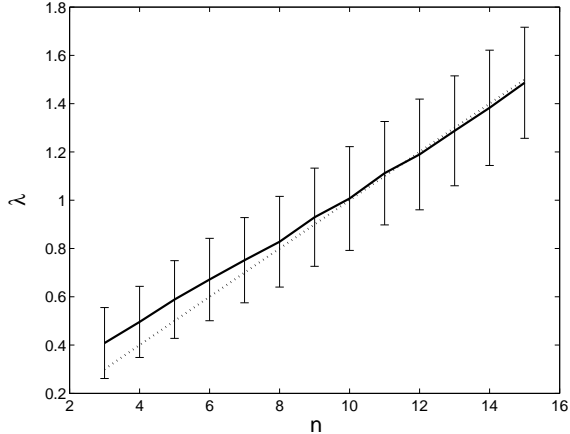}
     \caption{Largest eigenvalue of matrices of different sizes. Samples of 1000 matrices
     for each $n = 3$ through $15$
     were generated by choosing the matrix elements randomly from a distribution
     with average $\mu = 0.15$ and standard deviation $0.3$. The solid line connects the
     average value of $\LL$ for each sample, and
     the error bars are the standard deviations. The dotted line
     shows the analytic prediction $\langle \LL \rangle \sim n\mu$.}
    \label{f:ev_size_arb}
\end{figure}

The dependence of the largest eigenvector on matrix properties
relates the average convergence rate to parameters such as system
size and organizational structure. If the organization of the
problem solving process is flat, i.e., the level of interaction
between all tasks is arbitrary, most of the entries of the
interaction matrix are non-zero. In this case, the largest
eigenvalue of $\A_0 + \M$ is given on average by \cite{Juhasz}
\begin{equation}\label{avg_ev_arb}
\LL \sim n\mu + O(\sqrt{n}),
\end{equation}
where $n$ is the size of the matrix $\A_0 + \M$ and $\mu$ is the
average value of its elements. If we further assume that the
interactions are symmetric, that is, individual $i$'s results have
the same effect on individual $j$ that individual $j$'s have on
individual $i$, a more precise result is available
\cite{FurediKomlos}:
\begin{equation}\label{avg_ev_sym}
\LL = n\mu + \sigma^2/\mu.
\end{equation}
where $\sigma^2$ is the variance of the elements of $\A_0 + \M$.
The linear growth of $\LL$ with the number of independent tasks
($n$) and the tendency of interactions to hinder progress ($c$)
shows that the time to solution for a flat organizational
structure grows quickly as the project complexity increases. The
growth of $\LL$ with $n$ is demonstrated numerically in figure
\ref{f:ev_size_arb}.

In many situations, however, smaller teams work on parts of a
problem which are then put together at a higher organizational
level. From the point of view of our dynamical model, this means
that each task interacts only with the other tasks of the team or
their higher order aggregates. The ensuing interaction matrix will
then have blocks of nonzero elements and zero entries elsewhere.
In this case, a slower growth of the largest eigenvalue with size
is obtained \cite{stabilityofecosystems}:
\begin{equation}\label{avg_ev_tree}
  \LL \sim \mu \ln n,
\end{equation}
implying that much larger groups can be stable when they are
structured in highly clustered fashion, i.e. teams.

\section{The effect of fluctuations} \label{s:fluct}
We now examine the effect of fluctuations on the dynamics of
cooperative problem solving. As discussed above, these
fluctuations are a key ingredient of any cooperative process and
are due to many different factors. Recall that the fluctuations
are included in the model by modifying the interaction matrix from
the simple unperturbed WTM to the matrix $\A_t = \A_0 + \M +
\ess_t$, where $\M$ is the matrix of the mean values of the
fluctuations and $\ess_t$ is a matrix of mean-zero random
variables representing their temporal variation.

In the absence of fluctuations, the process is governed entirely
by the unperturbed matrix $\A_0$. The dynamics are thus equivalent
to the average dynamics described in section \ref{s:avg_dyn}, with
the interaction matrix $\A_0$ instead of $\A_0 + \M$. That is, the
process either converges smoothly to solution, if the largest
eigenvalue of $\A_0$ is less than 1, or diverges smoothly away
from solution if this eigenvalue is greater than 1. Moreover, for
convergent processes, the magnitude of the largest eigenvalue of
$\A_0$ determines the time to solution; processes with larger
eigenvalues take longer to converge. This behavior was previously
discussed and demonstrated in figure \ref{f:avg_dyn}, above.

The fluctuations affect the dynamics of the process in two
distinct ways. First, if the fluctuations have a non-zero mean,
that is, they tend to help or hinder task completion, $\M$ is
non-zero and the average dynamics are different from the
unperturbed dynamics. Second, if the strength of the temporal
variation $\S_t$ of the fluctuations exceeds a certain threshold,
the dynamics of individual project realizations may deviate
substantially from the smooth evolution of the average dynamics,
possibly causing long delays. These two effects are explored in
sections \ref{s:shift_avg} and \ref{s:temporal_var}.

\subsection{Shifted average dynamics} \label{s:shift_avg}
As mentioned above, average values of the fluctuations have the
effect of shifting the average dynamics away from the unperturbed
dynamics.

The unperturbed system is given by $\x_{t+1} = \A_0 \x_t$, and the
convergence rate is determined by the largest eigenvalue $\LL_0$
of $\A_0$.  When fluctuations are added, the average evolves
according to
\[
\lb \x_{t+1} \rb = (\A_0 + \M) \lb \x_t \rb
\]
and the average convergence rate is shifted from $\LL_0$ to $\LL$,
the largest eigenvalue of $\A_0 + \M$. In this section for
simplicity we will proceed using the discrete time formulation
only; the results which follow are applicable in the continuous
model as well.

Myriad forms are possible for the average fluctuation matrix $\M$.
A reasonable form to consider is that where each element of $\M$
is proportional to the corresponding element of $\A_0$ by some
factor $m$. That is, if the $(2,3)$ element of $\A_0$ equals 0.4,
indicating a strongly hindering correlation, we would have $M_{23}
= 0.4m$, while for a weakly hindering element such as $(A_0)_{12}
= 0.03$ we would take $M_{12} = 0.03m$. This assumes that the
relative strength of the correlations is not affected on average
by the fluctuations. One can thus model a wide variety of behavior
by varying $m$, even to have negative values. However, the most
reasonable values for $m$ are positive, since unforeseen events
typically hinder rather than help the completion of the project on
average.

The average interaction matrix thus takes the form
\[
\A_0 + \M = (1 + m)\A_0
\]
and its largest eigenvalue is given simply by
\[
\LL = (1 + m)\LL_0.
\]
This value determines the average dynamics of the perturbed
system, as described in section \ref{s:avg_dyn}. In particular,
for positive values of $m$, the average effect of the fluctuations
is to increase the time to solution. Moreover, if $\LL_0 < 1$ but
\[
m > \frac{1}{\LL_0} - 1,
\]
the fluctuations cause the project to diverge on average even
though an unperturbed treatment would predict convergence. This is
a far more likely scenario the closer $\LL_0$ is to 1, i.e., the
slower the predicted convergence in the unperturbed system.

For an example, please consider the average dynamics shown in
figure \ref{f:avg_dyn}, above. The curves in this plot were
generated by varying the off-diagonal entries of $\A_0$ from table
\ref{t:WTM}, above, taking $\M = \mathbf{0}$. Note that a similar
plot would be obtained by keeping $\A_0$ constant and varying $m$.
Assuming that the dashed line with $\LL = 0.87$ corresponds to
$m=0$, the solid line could be generated by taking $m = 0.103$ and
the dotted line by taking $m = 0.207$. In fact, given any
fluctuation in this system with $m > 0.149$, the dynamics of the
perturbed system will be divergent although the unperturbed system
converges. Moreover, even small $m$ can cause a huge increase in
convergence time as demonstrated by the dashed and solid curves in
the figure.

Generally speaking, even if $\M$ does not take the special form
studied above, a majority of positive entries in $\M$ indicate
that the average time to solution will very likely be larger than
the unperturbed prediction. It is a property of nonnegative
matrices that increases to individual elements results in an
increase of the largest eigenvalue \cite{Minc}. Even if $\A_0$ or
$\M$ has some negative entries, $\LL$ will still be greater than
$\LL_0$ on average according to \ref{avg_ev_arb} if the average
entry of $\M$ is positive.

\subsection{Temporal variability}\label{s:temporal_var}
The shift in largest eigenvalue brought about by $\M$ and
described in the previous section can engender delays in the
average dynamics, which are still described by either smooth
convergence to, or divergence away from, solution. Far more
interesting are the deviations from this smooth path brought about
by the temporal fluctuations which result in a jagged,
unpredictable evolution. As we now show, a high enough level of
temporal variation produces a skewed distribution of finishing
times and can result in long delays.

Recall that the temporal variation of the fluctuations is
represented by the matrix $\ess_t$, which is a matrix of random
variables with mean zero. To be as general as possible, we assume
that each entry of this matrix is independent of any other.
Analogously to the previous section, we also assume for
definiteness that the typical size of the fluctuations is
proportional to the respective average interaction values. For
example, if the element $(A_0 + M)_{35}$ = 0.1, then the standard
deviation of the (3,5) element of $\ess_t$ for any $t$ is equal to
$0.1s$, where $s$ is a constant of proportionality. These
statements can be concisely expressed by the correlation rule
\begin{equation}\label{correlation_rule}
  \lb (S_t)_{ij}(S_{t'})_{i'j'} \rb = s^2 [(A_0 + M)_{ij}]^2 \delta_{tt'} \delta_{ii'}
  \delta_{jj'},
\end{equation}
where $\delta$ is the Kronecker delta.

Notice that the state of the system is determined by many random
matrices multiplied together. This type of stochastic effect is
known as multiplicative noise. One aspect of systems with
multiplicative noise is that their distribution are approximately
log-normal, with parameters proportional to time. For small
fluctuations, the variance decreases with $t$ and the log-normal
distribution is very similar to a normal distribution, in which
events far from the average are very unlikely. However, once the
fluctuations exceed a certain threshold, the variance increases
with $t$ and the system's log-normal distribution develops a heavy
tail at long times, implying that events far from the average are
likely.
\begin{figure}[h!]
     %\centering
     \includegraphics[width=3in]{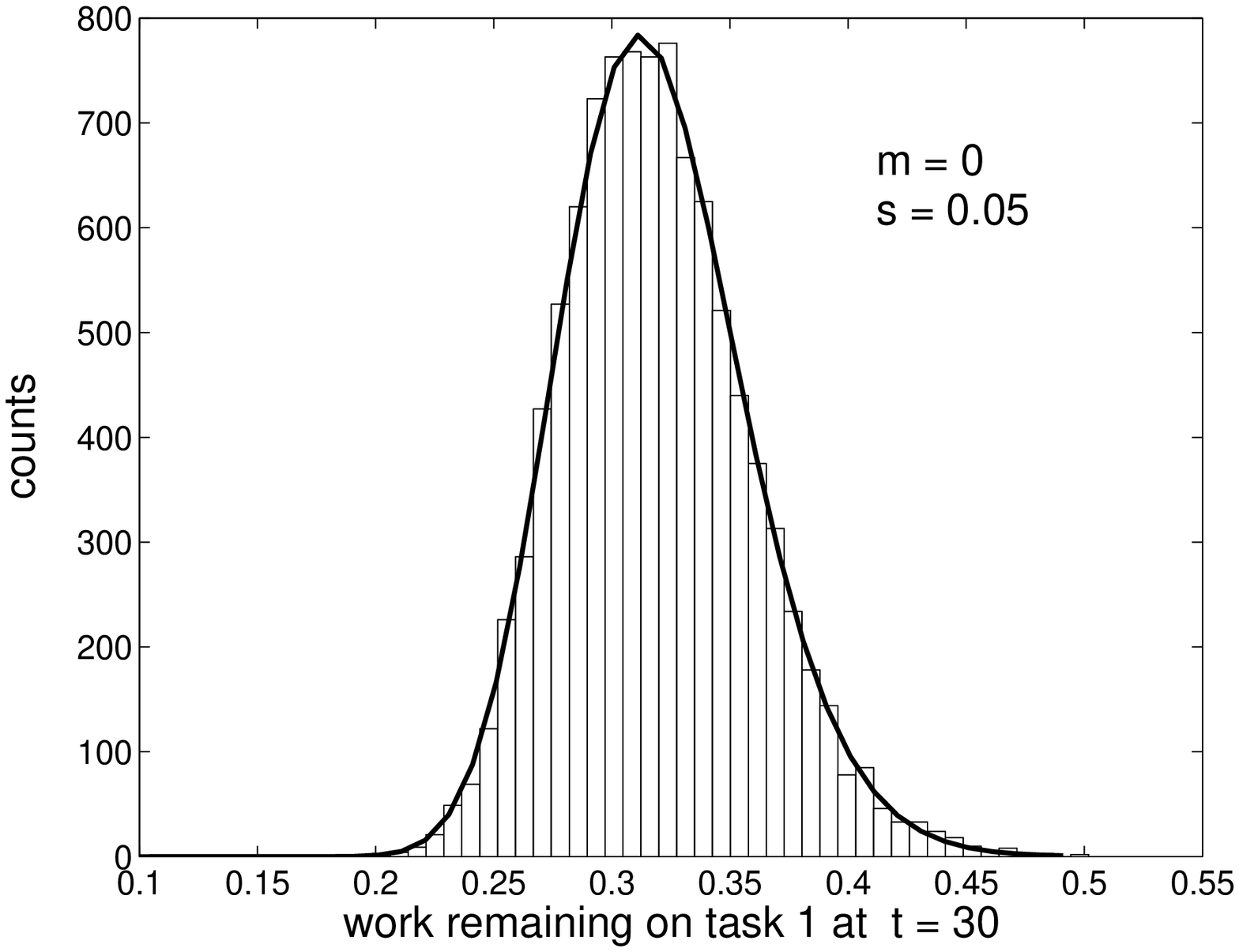}
     \includegraphics[width=3in]{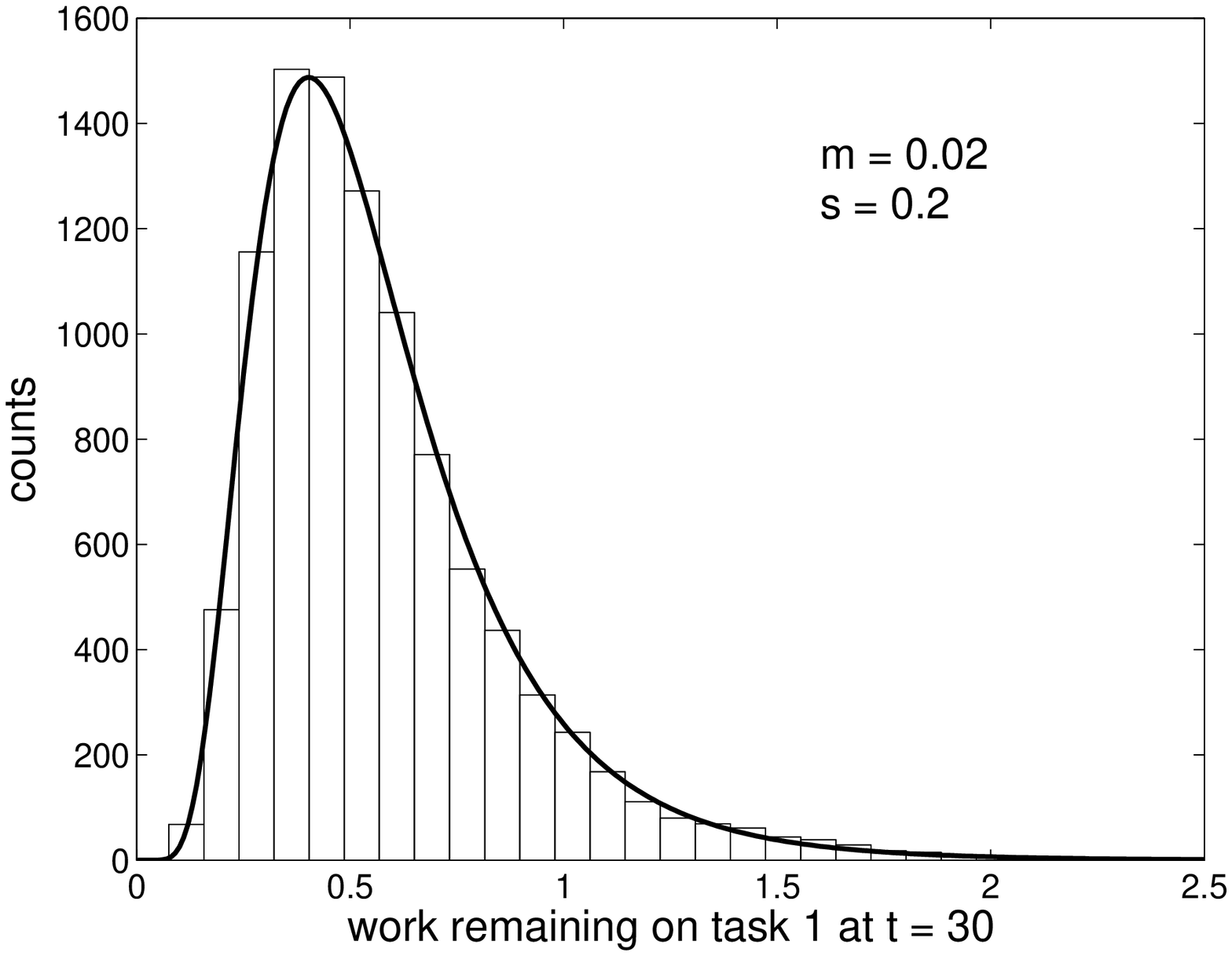}
     \centering
     \includegraphics[width=3in]{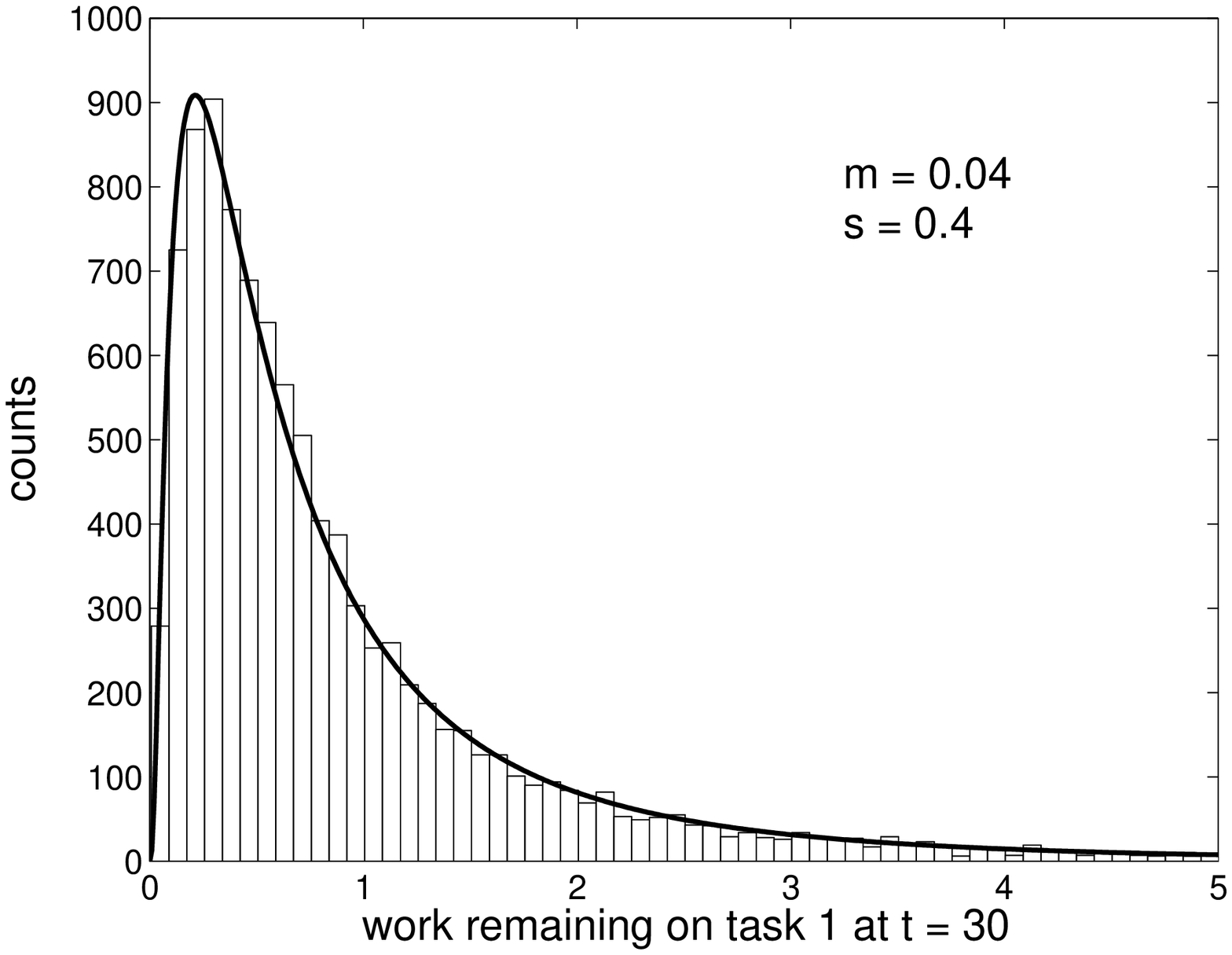}
     \caption{Histograms showing the distribution of the first component of $\x$ at time $t =
     30$ over 10,000 simulations of the vehicle interior appearance design
     project for various fluctuation strengths. The
     unperturbed WTM for this process is given in table \ref{t:WTM}.
     Recall the notation $m$ for the size of the fluctuation mean and
     $s$ for the strength of the temporal variation. The superimposed curves
     are log-normal probability functions calculated using the sample mean
     and variance and normalized by the bin width and height.
     The curve parameters are $\hat{\mu} = 0.316, 0.51, 0.603$ and
     $\hat{\sigma} = 0.123, 0.486, 1.019$ respectively by increasing fluctuation
     strength.}
    \label{f:x_hist}
\end{figure}

\begin{figure}[h!]
     \centering
     \includegraphics[width=3in]{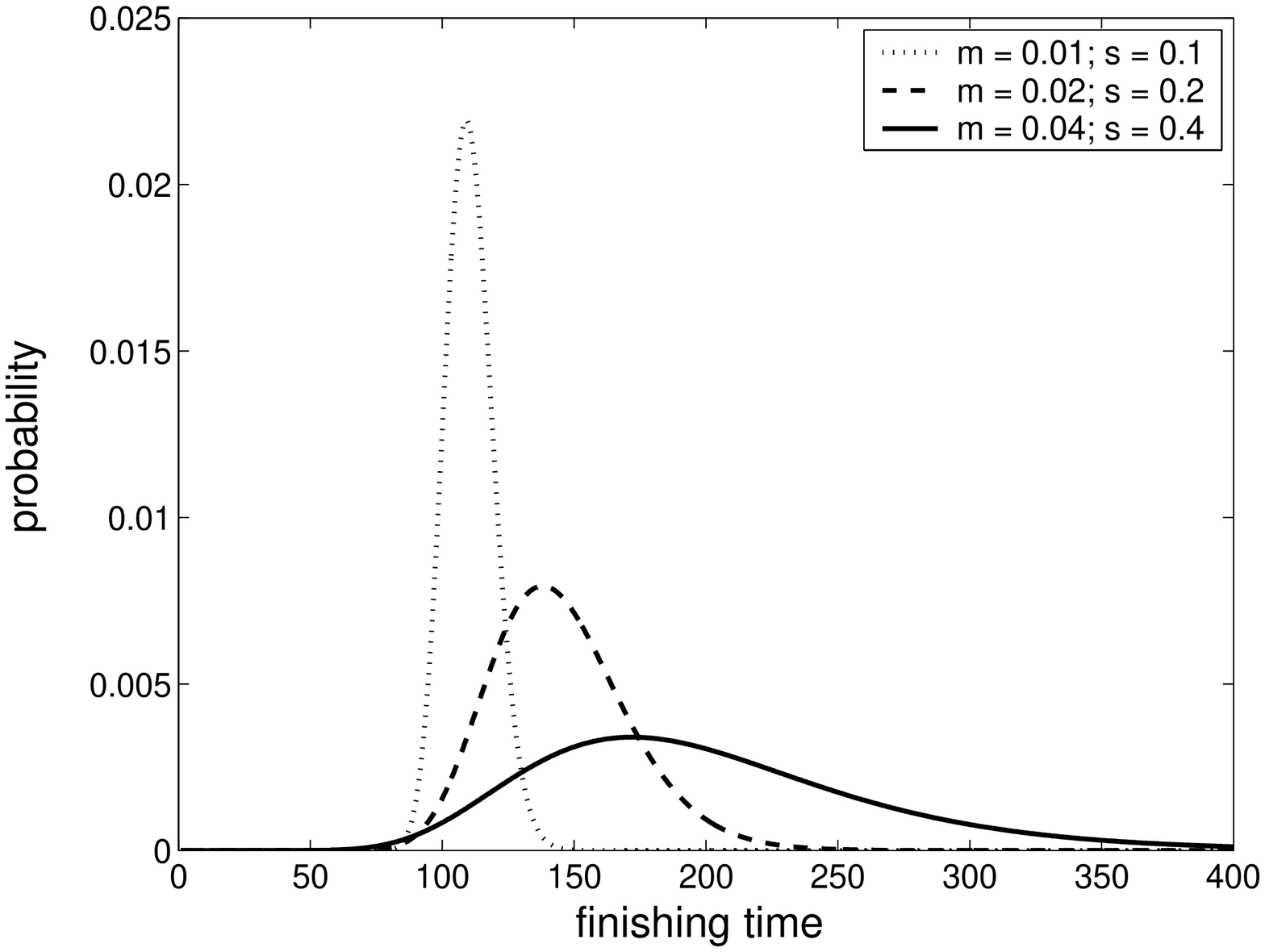}
     \caption{Distribution of finishing times for various
     fluctuation strengths over 10,000 simulations of the
     vehicle interior appearance design project. The curve parameters
     are $\hat{\mu} = 109.9, 142.61, 191.07$ and
     $\hat{\sigma} = .0831, .179, .3238$ respectively for
     increasing fluctuation strength.}
    \label{f:time_hist}
\end{figure}

The log-normal character of the system is shown in figures
\ref{f:x_hist} and \ref{f:time_hist}, below. In figure
\ref{f:x_hist}, a histogram showing the work remaining on task 1
in the vehicle interior appearance design project (see tables
\ref{t:DSM} and \ref{t:WTM}, above) is shown at $t = 30$ for
various fluctuation strengths. Each plot's distribution is
approximately log-normal as shown in the figure by the fitted
curves; note the differences in shape as the fluctuation strength
increases. In figure \ref{f:time_hist}, the time to solution is
plotted for several levels of noise; the stopping criterion is
that all components of the state vector be smaller than 0.05. This
log-normal distribution of solution times is in agreement with
previous results on cooperative problem solving
\cite{Clearwater_science}.

In order to understand the conditions under which fluctuations
lead to large deviations from the average dynamics, we study the
moments of the state vector. In general, the dynamics of systems
whose second or third\footnote{Large higher-order moments also
indicate deviations from the average but with smaller
probability.} moments $\lb |\x_t|^2 \rb$ and $\lb |\x_t|^3 \rb$
are significantly larger than the powers of the average $|\lb \x_t
\rb |^2$ and $| \lb \x_t \rb |^3$ are dominated by large
deviations from the average.

In our particular case, the moments may become orders of magnitude
larger than powers of the average as time goes on. This is because
the moments can diverge in time even as the average converges to
zero. In this situation the problem solving process is highly
susceptible to long delays brought about by large deviations from
the average. An example of the deviations brought about by small
and large fluctuations is shown in figure \ref{f:moment_dyn}
below.
\begin{figure}[h!]
     %\centering
     \includegraphics[width=2.75in]{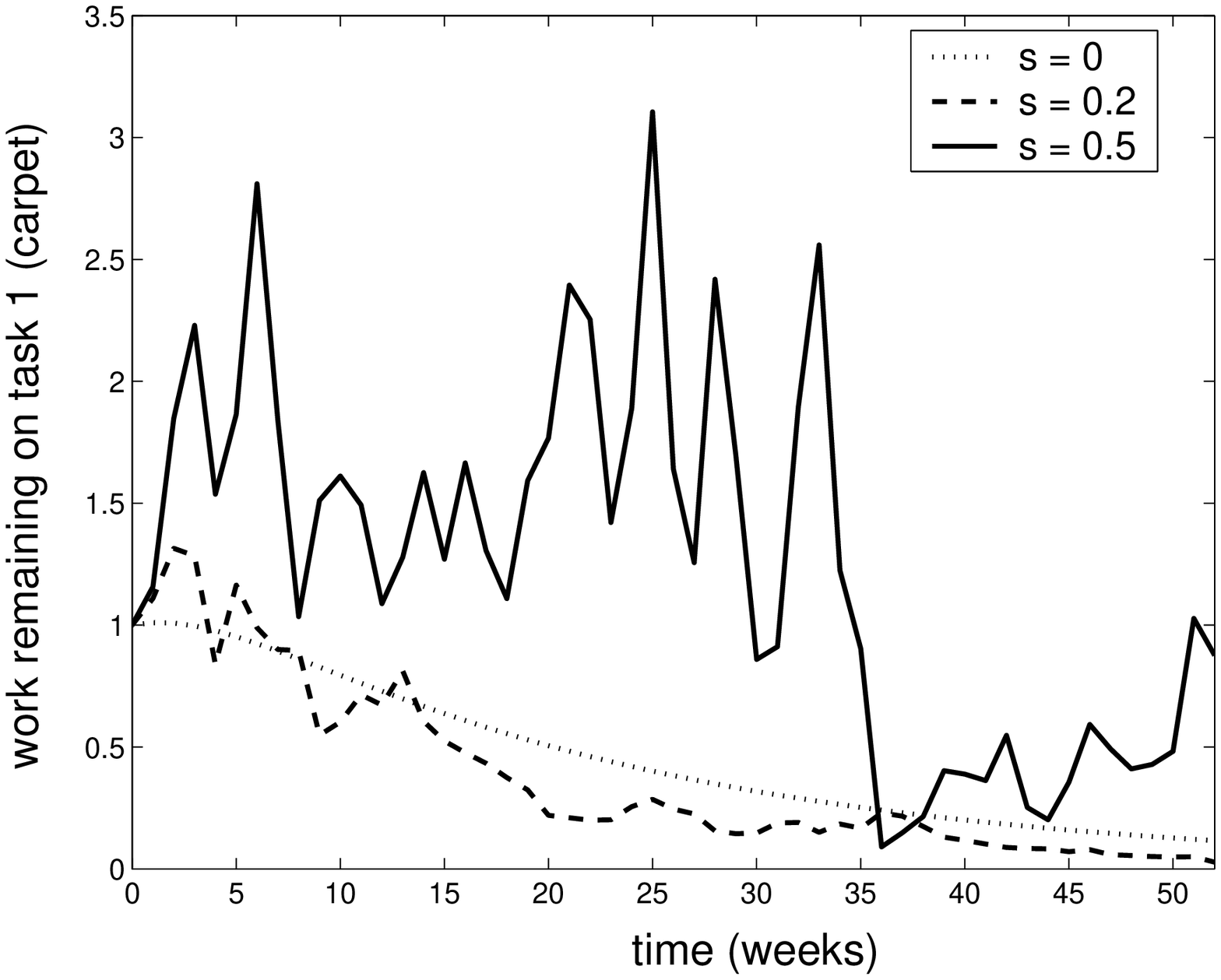}
     \includegraphics[width=2.75in]{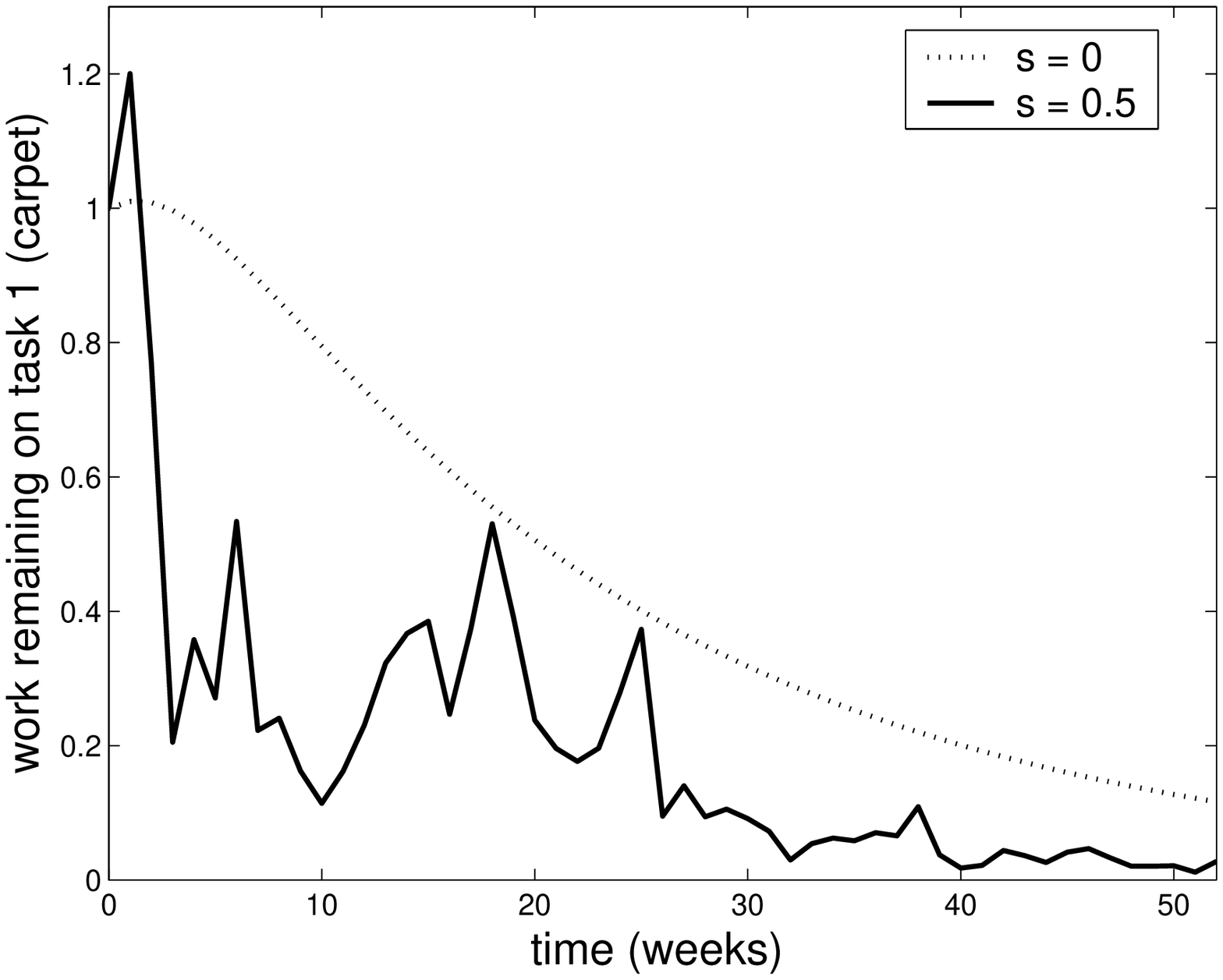}
     \caption{Example of dynamics for various fluctuation strengths.
     At left, the fluctuation causes a delay in task 1. At right,
     task 1 proceeds more quickly than an the unperturbed treatment
     would indicate.
     Recall that the parameter $s$ from (\ref{correlation_rule})
     measures the strength of the fluctuations.}
    \label{f:moment_dyn}
\end{figure}

We can obtain a mathematical condition on the matrices involved
which determines  when the deviations described above will occur.
To do this, recall expressions (\ref{avg_dyn_asympt}) for the
evolution of the average state, which imply that powers of the
average are given by
\begin{equation} \label{powers_of_avg}
\begin{array}{lllll}
|\lb \x_t \rb|^p &=& c^p e^{t p\ln \LL}&,& \mbox{discrete} \\
|\lb \x_t \rb|^p &=& c^p e^{t p(\LL-1)}&,& \mbox{continuous}.
\end{array}
\end{equation}
for large $t$. Here we are using the notation $\LL$ for the
largest eigenvalue of $\A_0 + \M$ and we have normalized the right
eigenvector $\uu$ to have length 1, as is customary. In general,
the moment evolution at large $t$ can also be expressed in this
form:
\begin{equation}\label{moment_general}
  \lb |\x_t|^p \rb = c^p e^{tL_p}
\end{equation}
(or $\lb |\x(t)|^p \rb$ for the continuous case) where $L_p$ is
the $p$th moment Lyapunov exponent and is always greater than the
corresponding exponent in (\ref{powers_of_avg}).  The interesting
case described above occurs when $\LL < 1$, so that the average
dynamics are convergent, but $L_p > 1$ for $p = 2$ or 3 so that
the second or third moments diverge.

The Lyapunov exponents are generally difficult to calculate.
However, \cite{Wilkinson} gives a measure of the difference
between the Lyapunov exponents and the corresponding exponents in
the powers of the average:
\begin{equation}\label{epsilon}
  \varepsilon^2 = \frac{s^2}{\LL^2}\sum_{i,j}v_i^2[(A_0 + M)_{ij}]^2u_j^2.
\end{equation}
Large values of $\varepsilon^2$ indicate that $L_p$ is
significantly larger than $p \ln \LL$. In this case, even a
process which converges rapidly on average, i.e. with a low $\LL$,
may undergo large deviations and encounter delays. When
$\varepsilon^2$ is small, deviations are unlikely unless the
convergence of the average is very slow, i.e. $\LL$ is close to 1.
Since the variance of the $(i,j)$ element of the fluctuations is
given by $s^2 [A_0 + M)_{ij}]^2$, it is clear from equation
(\ref{epsilon}) that the effect of fluctuations is increased as
their magnitude increases, as shown in figure \ref{f:moment_dyn}.

To show analytically how $\varepsilon^2$ affects the dynamics, we
consider cases in which the Lyapunov exponent in equation
(\ref{moment_general}) can be approximated calculated. When the
evolution is discrete and $\varepsilon^2$ small, the moments
evolve according to \cite{Wilkinson}
\begin{equation}
\begin{array}{lllll}
 \lb |\x_t|^2 \rb &=& c^2 e^{t(2 \ln \LL + \varepsilon^2)}&,& \mbox{second moment} \\
 \lb |\x_t|^3 \rb &=& c^3 e^{t(3 \ln \LL + 3\varepsilon^2)}&,& \mbox{third moment}.
 \end{array}
\end{equation}
These expressions show that the moments will diverge as
$\varepsilon^2$ increases.

\begin{figure}[h!]
     %\centering
      \includegraphics[width=2.5in]{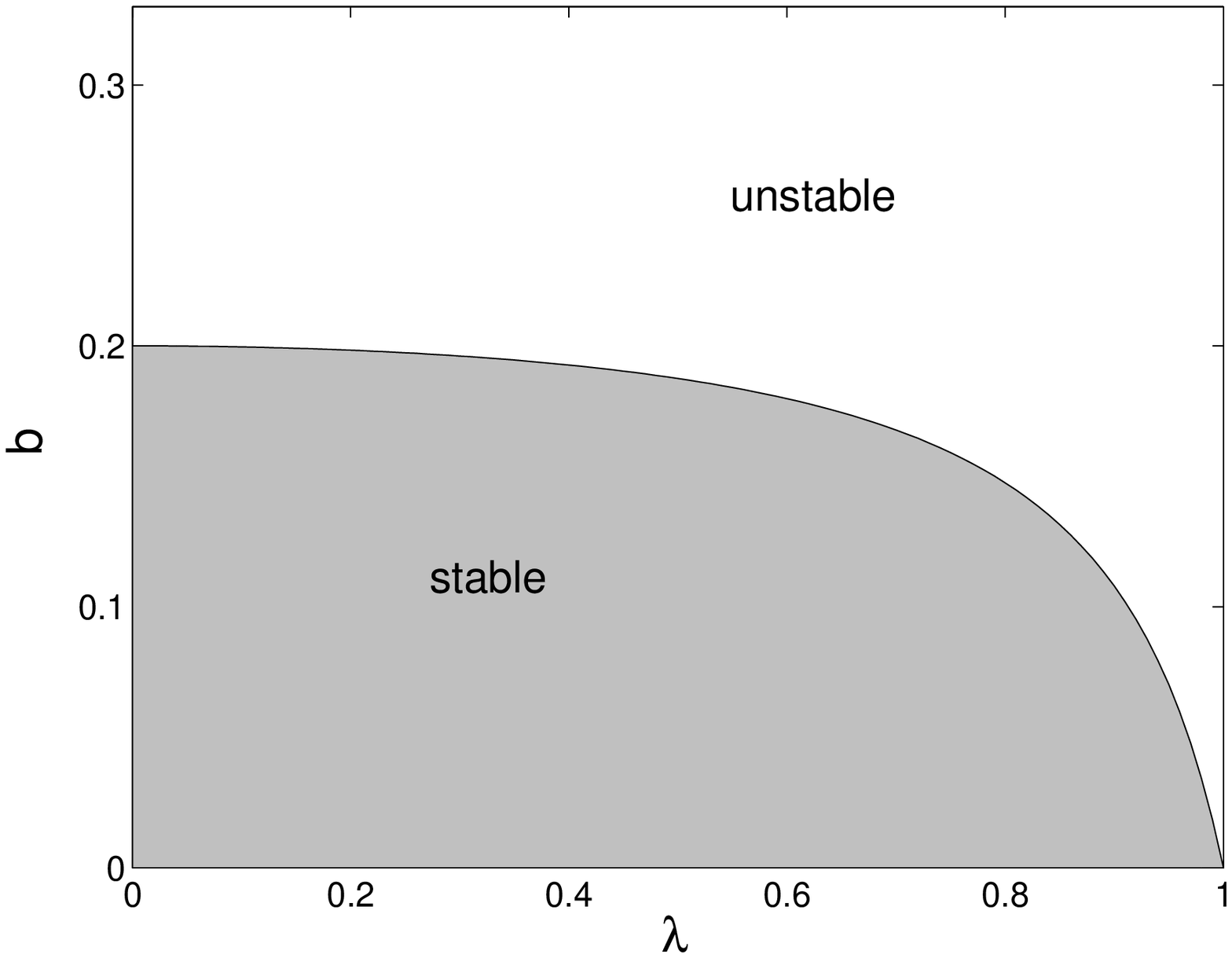}
      %\hspace{0.2in}
      \includegraphics[width=2.5in]{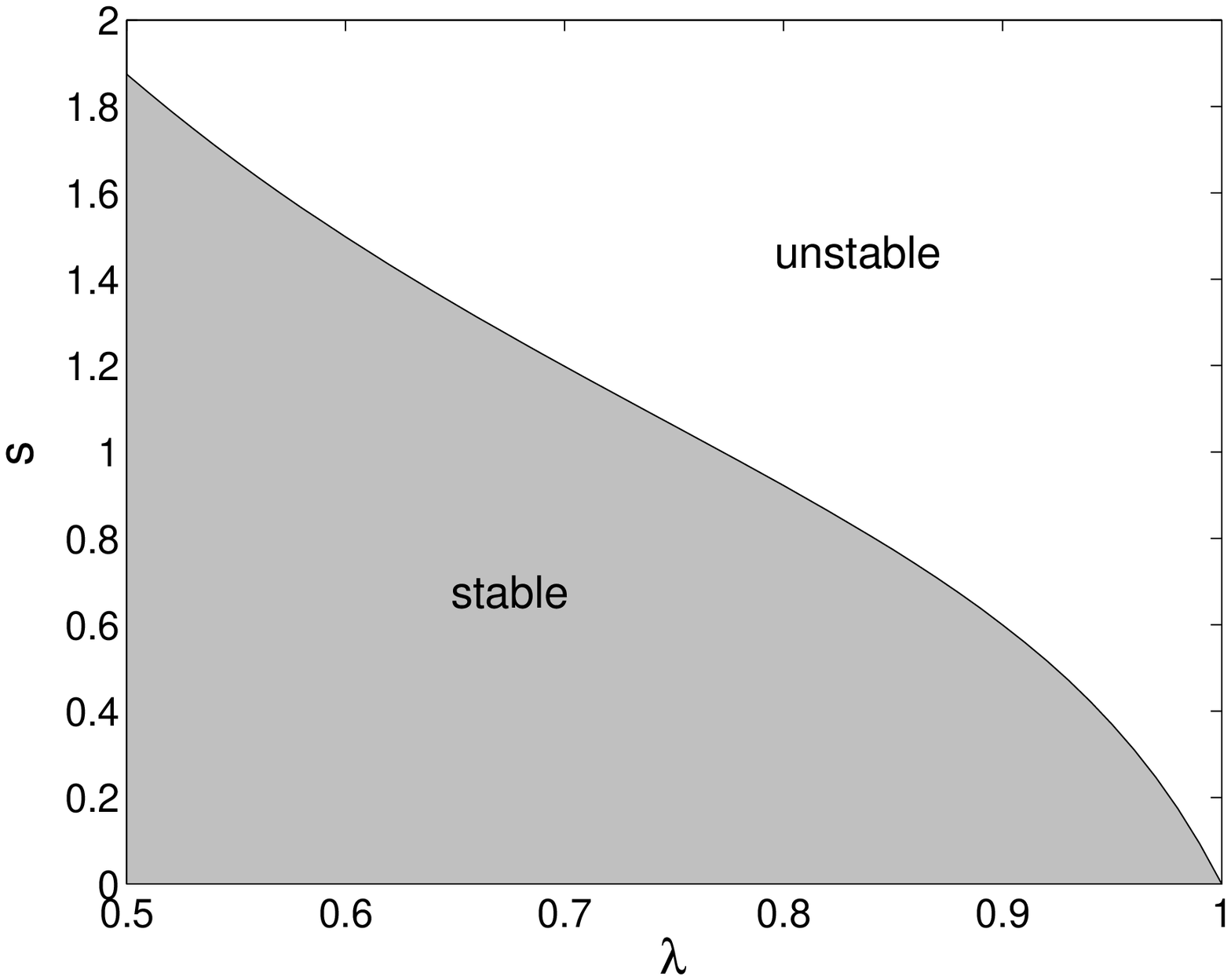}
     \caption{Stability diagram of a typical system with $n = 5$. Recall that $b$ is the
     absolute strength of the fluctuation, while $s$ is the strength proportional to the average size of the
     entries of $\A_0 + \M$.}
    \label{f:phaseplots}
\end{figure}

In the case in which the elements of the fluctuations all have the
same variance $b^2$, \cite{Wilkinson} provides a measure of the
critical strength of this variance above which deviations are
likely:
\begin{equation}\label{bcrit}
  b_c^2 = \frac{1}{n + \frac{v^2\LL^2}{1 - \LL^2}}.
\end{equation}
When the variance of the fluctuations exceeds $b_c^2$, the second
moment of the system will diverge. By applying the relation
(\ref{avg_ev_arb}) we can find a corresponding critical value
$s_c^2$, and thus determine the stability diagram of the project
dynamics. The critical values are plotted for various $n$ in
figure \ref{f:phaseplots}. When $\LL < 1$ but the fluctuation
strength is larger than $s_c^2$ or $b_c^2$, deviations from the
average are likely and although the process converges on average,
significant delays are likely to occur in particular realizations
of the project.

In the continuous case, analytic results are only available for a
two-dimensional system (two correlated tasks) and small noise. In
this case, the moments evolve according to
\cite{Arnoldperturbation}
\[
 \lb |\x_t|^p \rb = c^p \exp\big(t[p(\LL - 1) + p^2\varepsilon^2/2 + p\gamma_1 + pO(\varepsilon^2) +
 O(p^2)]\big)
\]
where $\gamma_1$ is a complicated function of the noise and the
eigenvectors $\uu$ and $\vv$ and $O(\varepsilon^2)$ means an
unspecified expression whose magnitude is on the order of
$\varepsilon^2$.

\section{Conclusion} \label{s:conclusion}
In this paper we presented a dynamical theory of  cooperative
problem solving projects consisting of any number of interrelated
tasks. The model incorporates the unavoidable fluctuating nature
of the interactions in such projects, which reflect the natural
variations in the amount of progress made on each task, as well as
the varying relevance of information from the partial solutions of
each task to any other. While simple, the theory captures key
aspects of the dynamics, such as the diminishing value of
interactions as the tasks near completion, as well as allowing for
a satisficing decision of when the project is completed. It also
captures the fact that projects with too many tasks  do not
converge to a desirable solution at all.

By analyzing the stochastic dynamics we showed that as the
strength of the fluctuations in task effectiveness increases past
a certain threshold, the convergence of a typical process to a
solution is very likely to deviate from what an average treatment
would indicate. In order words, single problem solving instances
are likely to deviate significantly from what one would expect if
fluctuations were ignored. In particular, a cooperative problem
solving process may be pushed far from its goal, resulting in a
significant delay in the time to solution. While this effect is
most severe in problems whose average convergence is slow, large
fluctuations can also disrupt the smooth convergence towards a
solution in any project. This prediction is in agreement with
numerous empirical results.

Moreover, over appropriate time intervals the distribution of
performance among tasks is log-normally distributed, as shown in
previous work on cooperative problem solving efforts. The
appearance of a heavy tailed distribution above a threshold value
in task variability implies that repeated observations of the
evolution of a project will tend to markedly differ from the
average dynamics that other models have analyzed. In other words
while there might be instances where time to completion seems to
be in agreement with an estimate of the average lifetime of the
project, on other occasions project duration will be dramatically
longer, with the concomitant aggravation of not being able to
predict when such long delays will take place.

These results suggest that in order to have a predictable outcome
in large cooperative projects, effort should be made to organize
cooperative tasks in a way that minimizes both the instabilities
implied by large unstructured groups, and the fluctuating nature
of their contributions. This can be achieved by structuring tasks
in such a modular or hierarchical  way  so that  they effectively
interact with very few others. Furthermore, given the stochastic
nature of such processes,  one possible mitigating design factor
is to keep fluctuations in performance constrained to variations
far below their threshold value.

Given the importance of large teams in the design and solution of
technical problems and the increasing need for production agility,
the study of the dynamics of group problem solving offers new
insights into the organization of teams and their efficiency at
solving complex tasks.

\subsubsection*{Acknowledgement} This work was partially supported
by  the National Science Foundation under Grant No. 9986651
%%%%%%%%%%%%%%%%%%%%%%%%%%%%%%%%%%%%%%%%%%%%%%%%%%%%%%%%%%%%%%%%
\newpage
\section*{Appendices}
\appendix
%\section{Project Reorganization}  \label{s:reorginazation}
%The situation where a project undergoes a major reorganization can
%be easily treated using the formalism introduced in this paper. By
%major reorganization we mean a long-term change which cannot be
%accounted for by fluctuations alone. The reorganization is
%characterized by a change in the unperturbed WTM, as well as the
%fluctuation parameters $m$ and $s$.
%To treat this situation we simply model the first part of the
%project using the initial unperturbed WTM and fluctuation
%parameters, and then modify these values for the second half of
%the project. Doing this does not introduce any new phenomena into
%the dynamics beyond what has already discussed in this paper.
\section{Fluctuations have no effect on the average}
\label{s:avg_fluct_noeffect} As mentioned above, the average
dynamics of the system are not affected by the fluctuations. In
order to see this, notice that in the discrete case the system
evolves according to $\x_t =(\C + \B_t)\x_{t-1}$. Since the
fluctuations are Markovian, i.e. the components of the matrix
$B_t$ are independent of the state at previous times,
\[
 \lb\x_t\rb = [\C + \lb\B_t\rb ] \lb \x_{t-1} \rb
\]
and we are taking the expected value of $\B_t$ entry by entry,
$\lb (B_\tau)_{ij} \rb = 0$ for all $\tau$, $i$ and $j$, we have
$\lb\x_t\rb = \C\lb\x_{t-1}\rb$ and, extrapolating back to $t=0$,
we obtain the discrete version of equation (\ref{avg_dyn}).
Similarly, in the continuous case formal integration gives
\[
\x(t) = \int_0^t (\C-\I) \x(s)ds + \int_0^t d\W  \x(s).
\]
Since $d\W = \B(t) dt$ and the elements of $\B$ have mean 0,
\[
\left\lb  \int d\W  \mbox{\boldmath$\phi$}(\x,s) \right\rb = 0
\]
for any function $\mbox{\boldmath$\phi$}(\x,s)$ and the expected
value is given simply by
\[
\lb \x(t) \rb = \int_0^t (\A-\I) \lb \x(s) \rb ds,
\]
which is equivalent to the continuous version of equation
(\ref{avg_dyn}).
\section{Dynamics of a simple linear system} \label{s:lin_dyn}
For any dynamical system specified by the equations
\[
  \begin{array}{cclll}
\lb \x_{t+1} \rb &=& \C \lb \x_t \rb & , & \mbox{discrete} \\
d\lb \x \rb &=& \C \lb \x(t) \rb dt & , & \mbox{continuous}
\end{array}
\]
the state at any time may be expressed in terms of the initial
state by writing
\begin{equation}\label{avg_solution}
  \begin{array}{cclll}
\lb \x_t \rb &=& \C^t \x_0 & , & \mbox{discrete} \\
\lb \x(t) \rb &=& e^{(\C-\I)t} \x(0) & , & \mbox{continuous}.
\end{array}
\end{equation}
In the long time limit, the matrix product $\C^t$ and the matrix
exponential $e^{(\C-\I) t}$ in these equations are dominated by
the contribution from the largest eigenvalue for large $t$. To see
this, we diagonalize\footnote{It is not necessary to consider
defective (non-diagonalizable) matrices in our case because it is
impossible to exactly determine the values of the elements of the
interaction matrices.} $\C$ into the form $\C = \PP\Lmat
\PP^{-1}$, where $\Lmat$ is the diagonal matrix of eigenvalues of
$\C$. This simplifies equations (\ref{avg_solution}) because
\begin{eqnarray*}
\C^t &=& \PP \Lmat^t \PP^{-1} \\
&=& \PP \left( \begin{array}{cccc}
\lambda_1^t & & & \\
& \lambda_2^t & & \\
& & \ddots & \\
& & & \lambda_n^t
\end{array} \right) \PP^{-1},
\end{eqnarray*}
where the $\{\lambda_i\}$ are the eigenvalues of $\C$, and
\begin{eqnarray*}
e^{(\C - \I)t} &=& \PP e^{{\mbox{{\scriptsize \boldmath$\Lambda$}}} - \I t} \PP^{-1} \\
&=& \PP \left( \begin{array}{cccc}
e^{(\lambda_1-1)t} & & & \\
& e^{(\lambda_2-1)t} & & \\
& & \ddots & \\
& & & e^{(\lambda_n-1) t}
\end{array} \right) \PP^{-1}.
\end{eqnarray*}
Since $\LL$ is the largest eigenvalue, when $t$ is large $\LL^t$
and $e^{(\LL-1) t}$ are much, much greater than all other terms in
the above matrices. The above equations are thus dominated by
$\LL$ and its eigenvectors $\uu$ (corresponding column of $\PP$)
and $\vv$ (corresponding row of $\PP^{-1}$). Expanding the matrix
product with all entries of $\Lmat$ neglected except the ones with
$\LL$, we obtain (\ref{avg_dyn_asympt}).
\section{Log-normal distribution}\label{s:log_normal_dist}
The log-normal probability distribution describes a random
variable whose logarithm is normally distributed. The probability
density function is
\begin{equation}\label{log-normal}
  P(x) = \frac{1}{x\sigma\sqrt{2\pi}}e^{\frac{(\ln x - \ln
  \mu)^2}{2\sigma^2}}
\end{equation}
where $\mu$ and $\sigma$ are the mean and standard deviation of
the normal distribution that describes $\ln x$. The $p$th moment
of the log-normal distribution is given by
\begin{equation}\label{ln_moments}
  \lb x^p \rb = \exp \big( p\ln \mu + \frac{p^2\sigma^2}{2} \big).
\end{equation}
In our application, $\sigma^2$ and $\ln \mu$ are proportional to
$t$. For large $\sigma^2$ the moments diverge for large enough
$p$, causing large deviations from the average.
% ----------------------------------------------------------------
\newpage
\bibliographystyle{siam}
\footnotesize
\bibliography{hub_wilk}
\end{document}